\documentclass[12pt]{article}

\usepackage{amssymb,amsmath,mathrsfs,epstopdf,slashed,color}

\usepackage{ifpdf}
\ifpdf
\usepackage{graphicx}
\usepackage{hyperref}
\usepackage{cite}
\else
\usepackage[dvipdfmx]{graphicx}
\usepackage[dvipdfmx]{hyperref}
\usepackage[numbers]{natbib}
\usepackage{hypernat}
\fi

\allowdisplaybreaks[1]

\makeatletter

\@addtoreset{equation}{section}
\makeatother

\title{\bf Comments on String Theory Inspired Inflation and Cosmic Strings}
\author{S.-H. Henry Tye, and Sam S.C. Wong}

\begin{document}

\begin{titlepage}

\setcounter{page}{0}
  
\begin{flushright}
 \small
 \normalsize
\end{flushright}

\vskip 3cm

\begin{center}

{\Large \bf Helical Inflation and Cosmic Strings}  

\vskip 2cm
  
{\large S.-H. Henry Tye${}^{1,2}$, and Sam S.C. Wong${}^1$}
 
 \vskip 0.6cm

 ${}^1$ Institute for Advanced Study, Hong Kong University of Science and Technology, Hong Kong\\
 ${}^2$ Laboratory for Elementary-Particle Physics, Cornell University, Ithaca, NY 14853, USA

 \vskip 0.4cm

Email: \href{mailto:astye@ust.hk, scswong@ust.hk}{iastye at ust.hk, scswong at ust.hk}

\vskip 1.0cm
  
\abstract{\normalsize   
Recent BICEP2 detection of low-multipole B-mode polarization anisotropy in the cosmic microwave background radiation supports the inflationary universe scenario and suggests a large inflaton field range. The latter feature can be achieved with axion fields in the framework of string theory. We present such a helical model which naturally becomes a model with a single cosine potential, and which in turn reduces to the (quadratic) chaotic inflation model in the super-Planckian limit. The slightly smaller tensor/scalar ratio $r$ of models of this type provides a signature of the periodic nature of an axion potential.  
We present a simple way to quantify this distinctive feature.
As axions are intimately related to strings/vortices and strings are ubiquitous in string theory, we explore the possibility that cosmic strings may be contributing to the  B-mode polarization anisotropy observed.

}

\vspace{1cm}
\begin{flushleft}
 \today
\end{flushleft}
 
\end{center}
\end{titlepage}

\setcounter{page}{1}
\setcounter{footnote}{0}

\tableofcontents

\parskip=5pt

\section{Introduction}

Observations of the Cosmic Microwave Background (CMB) radiation provides strong support for the inflationary universe scenario that explains the origin of the hot big bang beginning of the universe \cite{Guth:1980zm,Linde:1981mu,Albrecht:1982wi}. CMB temperature anisotropies measured by COBE, WMAP and Planck are in excellent agreement with predictions of the simplest inflationary models \cite{Bennett:1996ce,Bennett:2012zja,Ade:2013uln}, which also predict a scale-invariant spectrum of gravitational waves \cite{Starobinsky:1979ty} that came from the quantum fluctuation of space-time during inflation \cite{Crittenden:1993wm,Frewin:1993dq,Harari:1993nb}. The gravity waves generate primordial polarization, including the so-called B-mode pattern \cite{Kamionkowski:1996zd,Seljak:1996gy,Zaldarriaga:1996xe,Kamionkowski:1996ks}. In particular, the recent data from BICEP2 \cite{Ade:2014xna} is consistent with predictions of the chaotic inflation (the quadratic version) model \cite{Linde:1983gd}. On the other hand, the B-mode signal seen by BICEP2 can also contain small contributions from other potential sources such as cosmic strings. 

The primordial microwaves are linearly polarized, which may be decomposed into E-mode and B-mode. Quantum (scalar) fluctuation of the inflaton (the scalar field responsible for inflation) leads to E-mode polarization while the quantum fluctuation of space-time metric during the inflationary epoch leads to both E- and B-mode polarizations. If the recent BICEP2 observation of the B mode polarization in CMB is confirmed, it would imply the presence of the tensor mode quantum fluctuation and so gravity is quantized. The only known consistent perturbatively quantized gravity theory is the (super)string theory. If string theory is the theory of nature, it should be able to explain the inflationary universe. Although there are a number of explicit realizations of the inflationary universe scenarios in string theory, a typical range of the inflaton field $\phi$, i.e., the field range that $\phi$ evolved during the inflationary epoch, is $\Delta \phi < M_{pl}$, where $M_{pl}$ is the reduced Planck mass. This property simply follows from the compactification of the extra dimensions present in string theory \cite{Baumann:2014nda}. Following from the Lyth bound \cite{Lyth:1996im},
\begin{equation}
 \frac{\Delta \phi}{M_{pl}} \ge  N_{e} \sqrt{r/8}
\end{equation}
where $N_e$ is the number of e-folds of inflation and $r$ is the tensor/scalar ratio. Taking $60 \ge N_e \ge 40$, we find that typical values of $r$ satisfies $r < 0.005$ for $\Delta \phi < M_{pl}$. This is much smaller than $r \simeq 0.2$ observed by BICEP2 \cite{Ade:2014xna}. 

One way to obtain a relatively large $r$ value in string theory is to employ the axion fields for inflation, which has been explored under the name of natural inflation \cite{Freese:1990rb,Adams:1992bn,Randall:1995dj} or axion monodromy \cite{Silverstein:2008sg,McAllister:2008hb,Marchesano:2014mla}. A very good feature of axion-generated inflation is the presence of the (approximate) shift symmetry: $\phi \rightarrow \phi \, +$ constant, a property noted some years ago \cite{Adams:1992bn}. As the inflaton, this property can protect the axions from the so called $\eta$ problem, namely having too steep a potential.
Being angular (or phase) fields, axions can perform helical-like or similar motions to extend the field range, thus allowing a larger effective field range for the inflaton \cite{Kim:2004rp,Kaloper:2008fb}.
Since axions are ubiquitous in string theory, such scenarios should actually appear rather naturally. 

So it is not difficult to come up with models that can fit the existing data. Here we present a simple two-axion helical model. 
The model reduces to a single cosine potential, which in turn reduces to a model closely resembles the quadratic version of chaotic inflation. Following Ref\cite{Kallosh:2014vja}, we see that this feature is quite natural in a large class of axionic models in the supergravity framework. Because of the periodic nature of an axionic potential, this cosine model can have a slightly smaller value of $r$ than that from chaotic inflation. This deviation is quite distinctive of the periodic nature of the inflaton potential. 

Here we present a simple way to search for this periodic feature by quantifying its deviation from the $\phi^2$ chaotic inflation \cite{Linde:1983gd}, a model that fits the existing data reasonably well. All physical parameters such as the runnings of the power spectra indices can be expressed in terms of $r$ and the scalar power spectrum index $n_s$, including the parameter $\hat \Delta$, which measures the deviation of the cosine model from $\phi^2$ chaotic inflation,
$$ {\hat \Delta} = 16 \Delta =  r+ 4(n_s-1) $$ 
where $\phi^2$ chaotic inflation has $\hat \Delta=0$. We show that other quantities such as runnings of spectral indices have very simple dependences on $\hat \Delta$. We see that a downward shift of $r$ from its value in the $\phi^2$ model by as large as ${\hat \Delta}= - 0.03$ (or about $20\%$ of $r$) is possible. As the data improves, a negative value of  $\hat \Delta$ can provide a distinctive signature for a periodic axionic potential for inflation. 

However, when there are many axions, or axion-like fields, with a variety of plausible potentials, the possibilities may be quite numerous and so predictions may be somewhat imprecise. Here, we like to point out that the presence of axions would easily lead to cosmic strings (i.e., vortices, fundamental strings and D1-strings), which may provide a relatively clean signature of string theory scenarios for the inflationary universe. This is especially relevant if cosmic strings come in a variety of types and tensions and maybe even junctions. It so happens that cosmic strings will generate some B-mode polarization as well. This  provides the motivation to further explore cosmic strings along this direction.

Although the helical inflation model generates a $r$ value consistent with the BICEP2 result, it typically yields a slightly smaller value. This may leave room for B-mode contributions coming from cosmic strings. Recent analyses of the BICEP2 and POLARBEAR data \cite{Ade:2014afa} suggests that some cosmic string contribution to the B-mode polarization is possible in the fitting of the B-mode power spectrum \cite{Lizarraga:2014eaa,Moss:2014cra}. In any case, since the B-mode spectrum from cosmic strings \cite{Pogosian:2003mz} is different from the B-mode spectrum from inflation, better B-mode polarization data will either provide evidence of cosmic strings or put a tight bound on its contribution. 

This paper is organized as follows. In Sec. 2, we present the helical model. Since the scale of the inflaton potential is essentially at the the grand unified theory (GUT) scale,  it is possible that a phase transition may have taken place during inflation. We make some preliminary observations on this issue. Since the helical  model reduces to a single cosine potential, we suggest a simple way to pick out the key feature of a periodic potential by comparing it to $\phi^2$ chaotic inflation.  This analysis is presented in Sec. 3. In Sec. 4, we discuss some properties of cosmic strings in relation to the B-mode polarization. We give our conclusions in Sec. 5.

\section{Models}

We are interested in a particular potential for 2 axions of the form,
\begin{equation}
     V(\phi_1,\phi_2) = V_1+V_2=V_0\left\{ 1- \cos\left({\phi_1\over f_1}\right) +   A\left[1-\cos\left( {\phi_1 \over f_1'}-{\phi_2\over f_2} \right)  \right] \right\}. \label{poten}
\end{equation}
This is a model of natural inflation with two axions $\phi_1$ and $\phi_2$. Two axions potential was written down in \cite{Kim:2004rp} to produce large effective axion scale from sub-Planckian scales. The constant is introduced so that $V=0$ when it hits the minimum. Let us first introduce this potential from the supergravity perspective and then examine it as an inflation model to give some general properties. We believe there exist a model such that the axion directions play the central role of inflaton while their real partner do not change drastically during inflation. This requires further exploration of models and we focus on the axions only for this work. 

\subsection{Realization in Supergravity}

The supergravity scalar potential is given by
\begin{equation}
    V = e^{K} \left(K^{I \bar{J}} D_I W D_{\bar{J}} {\overline W} - 3W {\overline W}\right).
\end{equation}
Consider a superpotential of the form,
\begin{equation}
   W = W_0 + A_1e^{-a_1T_1} +A_2e^{-a_2T_1 - bT_2}. 
\end{equation}
This is a racetrack like model\cite{Escoda:2003fa,BlancoPillado:2004ns} and $T_1,T_2$ can be any 2 moduli fields. It is natural to have non perturbative terms for the moduli fields and these non perturbatuve terms are powerful for moduli stabilization in string theory\cite{Kachru:2003aw}. Due to the cross terms between $W$ and ${\overline W}$ or $ D_I W$ and $D_{\bar{J}} {\overline W}$, terms of these forms will show up in the scalar potential,

\begin{center}
  \begin{tabular}{c|c  c c}
                                                     &  $\;W_0\;$              &     $+ A_1e^{-a_1T_1}$                                    & $+A_2e^{-a_2T_1 - bT_2} $    \\  \hline
            $     W_0^*     $                  &                      &  $  W_0^*A_1  e^{-a_1 t_1}e^{-i a_1\tau_1}$   &  $  W_0^*A_2  e^{-a_2t_1-bt_2}e^{-i a_2 \tau_1-ib \tau_2}     $         \\
$   A_1^*e^{-a_1T_1^*}$     &         &       &      $A_1^*A_2  e^{-(a_1+a_2) t_1-bt_2}e^{-i (a_2-a_1) \tau_1-ib \tau_2}$   \\                  
   $   A_2^*e^{-a_2T_1^*-bT_2^*} $    &         &  &                     \\
  \end{tabular}
\end{center}
The diagonal terms are independent of $\tau_1,\; \tau_2 $ while the lower left triangle are the complex conjugate of the given terms. They sum up to produce the cosine potentials,
\begin{equation}
\label{3cos}
       \cos (a_1\tau_1), \quad \cos(a_2\tau_1 + b\tau_2), \quad \cos[(a_1-a_2)\tau_1 - b\tau_2].
\end{equation}
In general, all fields except those with relatively flat potentials will quickly reach their stabilized values. The fields with relatively flat potential would evolve over a period of time before they can reach their stabilized values. While they are evolving towards their stabilized values, inflation takes place. We shall assume that all other fields are stabilized except $\tau_1 \sim \phi_1$ and $\tau_2 \sim -\phi_2$. This does not rule out the existence of fields with degenerate vacua. Typically they are axions with their shift symmetry intact. We shall come back to them later. For $|W_0| >> |A_i|$, we may ignore the third term in Eq.(\ref{3cos}). 

The first 2 terms in Eq.(\ref{3cos})  are the desired  terms to produce the potential (\ref{poten}). The coefficients of these cosines and the additional terms depend on the other modulus fields, most of them are generically protected by moduli stabilization at much higher scale. 
At this stage, we just assume that they are stabilized in order to ignore their motion during inflation. Ref\cite{Kallosh:2014vja} provides a simple realization of this type of models.

\subsection{The Helical Model}

We will now show that how the helical model (\ref{poten}) can be similar to chaotic inflation with  consistent super-Planckian field excursion. We focus on the region ${\phi_1 \over f_1} < 1$ such that the first term of the potential is quadratic like, 
\begin{equation}
   V_1 =   V_0\left[ 1- \cos\left({\phi_1\over f_1}\right)\right] = {1\over2} {V_0\over f_1^2} \phi_1^2 + \dots
\end{equation}
The overall scale $V_0$ is determined by fitting the scale of inflation to be the GUT scale. For the second term of the potential, we choose $f_1' \ll f_1$. Now $ {f_1 \over 2\pi f_1'}$ measures the number of cycles that $\phi_2$ can travel for $0<\phi_1 < f_1$. Although $\phi_2$ has a shift symmetry $\phi_2 \rightarrow \phi_2 + 2\pi f_2$, the whole system do not return to the same configuration after $\phi_2$ has traveled for one period because $\phi_1$ has also moved. Thus, instead of a shift symmetry, the system has a helical symmetry.  This is in fact the idea of completing natural inflation \cite{Kim:2004rp} and axion monodromy\cite{McAllister:2008hb}, where the field range of $\phi_2$ can now be extended way above the scale of $f_2$ and even beyond the Planck scale. 
 \begin{center}
            \includegraphics[scale=0.4]{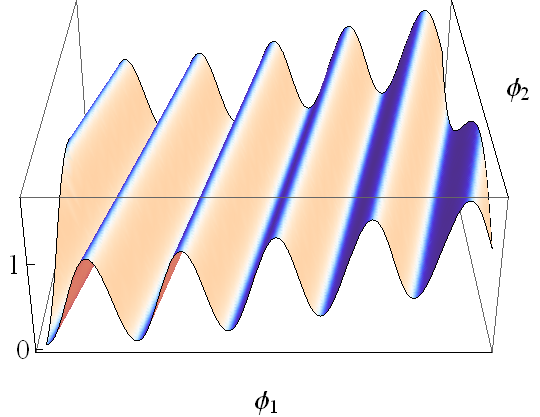} \\
            {\small{} FIG. 1, Potential of the model (\ref{poten}). The range of $\phi_2$ is a period $2\pi f_2$ and the oscillation period in $\phi_1$ is $2\pi f_1'$. The rising of oscillation in $\phi_1$-direction is due to $V_1$ while the path of the trough is produced by the second term in the potential. }
 \end{center}

As illustrated in FIG. 1, the bottom of the trough of the potential is the minimum of $V_2$, the second term in $V$ (\ref{poten}). This path of the trough is given by
\begin{equation}
\label{trough}
         {\phi_1 \over f_1'}-{\phi_2\over f_2} = 0 .
\end{equation}
For large enough $A$, this condition will be obeyed and
the inflaton can roll down along this path. Moving along the path of the trough (\ref{trough}), we see that the second term in the potential (\ref{poten}) vanishes and so the potential $V$ reduces to that with only the first term,
\begin{equation}
\label{1cos}
 V_1(\phi_2) \simeq V_0\left\{ 1- \cos\left({f_1' \phi_2\over f_1f_2}\right) \right\}
\end{equation}
where the range of the inflaton field $\phi_2$ can easily be super-Planckian.

To properly normalize the fields, one defines two normalized orthogonal directions,
\begin{equation}
    X = { f_1' \phi_1 + f_2 \phi_2 \over \sqrt{f_1'^2 + f_2^2}  },   \quad    Y= { f_2 \phi_1 - f_1' \phi_2 \over \sqrt{f_1'^2 + f_2^2}  },
\end{equation}
the inflaton will roll along the $X$-direction while $Y$-direction is a heavy mode which can be integrated out. Note that the system is insensitive to the magnitude of $A$ as long as it is greater than ${\cal O}(1) $ such that $Y$-direction is heavy enough. For instance, the slow roll parameters $\epsilon$ and $\eta$ are not affected which means that the observables $n_s$ and $r$ are insensitive to $A$. 
Since the $X$ path is already at the minimum of the second term, at $Y=0$, the effective potential along $X$ is determined by $V_1$,
\begin{equation}
\label{Xcosine}
    V(X) = V_0\left\{ 1- \cos\left({X\cos \theta /f_1}\right) \right\} \approx  {1\over2}  {V_0\over f_1^2} \cos^2 \theta X^2 , \quad \mbox{where  }  \cos^2{\theta} = {  f_1'^2  \over f_1'^2 + f_2 ^2}.
\end{equation}

As claimed, the helical model effectively reproduces the chaotic inflation potential,
\begin{equation}
\label{Xpot}
      V_{eff}(X) = {1\over 2} m^2 X^2, \quad   \mbox{where }m^2 =  {V_0\over f_1^2} {  f_1'^2  \over f_1'^2 + f_2 ^2}.
\end{equation}
As we know, the field excursion in chaotic inflation is $\Delta X \sim 14 M_{pl}$ and $X_{end} = \sqrt{2} M_{pl}$, where $M_{pl}=2.4 \times 10^{18}$ GeV is the reduced Planck mass. We can thus use this to constrain the model. 
\begin{equation}
    \Delta X = \Delta \phi_1 \sqrt{1+ \left({f_2 \over f_1'}\right)^2} \approx  {f_1 f_2 \over f_1' } >  14 M_{pl} ,
\end{equation}
where we have assumed that $f_2 \gg f_1' $ and $\Delta \phi_1 \approx f_1$. In this case, we have
$$X \simeq \phi_2 \quad \quad m^2 \simeq V_0/(\Delta X)^2 $$
so the model is essentially reduced to a two-parameter ($V_0$ and $\Delta X \simeq f_1f_2/f_1'$) model, and
 we have verified the second part of the claim, that the effective field range can be super-Planckian. 
For instance, $f_1=f_2 =M_{GUT}= 10^{-2}M_{pl}$ and $f_1' =10^{13}\mbox{GeV} =5\times 10^{-6}M_{pl}$ will give the field range of approximately $20M_{pl}$. In this case ${f_2 \over 2\pi f_1'}=318 $, so $\phi_2$ can rotate 300 more times than its period indicates. On the other hand, with the combination $ {f_1 f_2 \over f_1' }=20 $, the magnitude of $V_0$ is fixed to be $1.5 \times10^{-8} M_{pl}^4 $ by the scale of inflation. That is, we require $m^2 \simeq 4 \times 10^{-11} M_{pl}^2$. 

The difference to the $r$ value in the chaotic inflation model lies in the cosine form potential for $\phi_1$. 
We have numerically solved the equation of motion for the model (\ref{poten}) with cannonically normalized kinetic terms and evaluated the scalar spectral index $n_s = 1-6\epsilon +2\eta$ and tensor to scalar ratio $r=16\epsilon$. The slow roll parameters $\epsilon={M_{pl}^2 \over 2}\left({V'\over V}\right)^2 ,\; \eta=M_{pl}^2 \left( {V''\over V} \right)^2$ are defined along the inflaton direction and evaluated at $N_e$ e-folds before the end of inflation. 
We present a comparison between (\ref{poten}) and chaotic inflation,
\begin{center}
  \begin{tabular}{|c|c  c c |}\hline
       $N_e =60 -50 $    & $ n_s$   &  $r$    &   $ \hat{\Delta}=r+4(n_s-1)$  \\  \hline
       Chaotic Inflation    &    $0.967-0.960 $  &   $0.132 - 0.158$       &   $0$         \\
         Helical Model   (${ f_1 f_2 \over f_1' }=20 M_{pl}$)      &   $ 0.966-0.960  $  &   $0.124-0.150 $  &     $-0.011$          \\
  Helical Model   (${ f_1 f_2 \over f_1' }=14 M_{pl}$)      &   $ 0.966-0.960  $  &   $0.115-0.141 $         &       $-0.020$           \\
      Experimental Data               &     $ 0.9607 \pm 0.0063  $  &      $   0.16^{+0.06}_{-0.05} $          &                   \\ \hline
  \end{tabular}
\end{center}
\begin{center}
   {\small{} TABLE 1, Comparison of $n_s$ and $r$ between chaotic inflation and the helical model with ${ f_1 f_2 \over f_1' }=20 M_{pl}$ and ${ f_1 f_2 \over f_1' }=14 M_{pl}$ . Numbers going from left to right correspond to $N_e=60$ to $N_e=50$ respectively.  The $n_s$ value comes from Planck \cite{Ade:2013uln} while the $r$ value comes from BICEP2 \cite{Ade:2014xna} after subtracting the dust contribution. Note that BICEP2 reported $r=0.20^{+0.07}_{-0.05}$ before this subtraction. Note that the predicted value of $\hat \Delta=16 \Delta$ is independent of $N_e$.}
\end{center}

The helical model has typically a slightly smaller $r$ than chaotic inflation since $\cos (\phi_1/f_1)$ potential is less steep due to corrections from sub-leading order terms in $\cos (\phi_1/f_1)$, such as the $-{\phi_1^4}/{4!} $ term. 
For a fixed (or precisely measured) power spectrum index $n_s$, the number of $N_e$ is determined in both chaotic inflation and the helical model. The measurement of $r$ can then determine whether the feature of the periodic potential is present or not. This motivates us to do a more detailed analysis of the periodic potential in the next section.
 

\subsection{Phase Transition and Inflation}

The scale of the inflaton potential is given by the value $r$ and the amplitude $A_s= 2.2 \times 10^{-9}$ of the scalar power spectrum \cite{Ade:2013uln},
\begin{equation}
\label{GUTinf}
 r\simeq 0.2 \rightarrow  V_{infl}^{1/4}=\left( {3\pi^2\over 2} A_s r \right)^{1/4}M_{pl} \simeq 2 \times 10^{16} \mbox{GeV} 
 \end{equation}
This scale is very close to the grand unification (GUT) scale, so it is natural to ask whether it is possible that a phase transition, say the spontaneous symmetry breaking of the grand unified theory,  can take place during inflation ? If so, what is the impact on the inflationary universe scenario ? Here, let us do a preliminary investigation on this issue.

As is well known, the finite temperature effect would prevent the spontaneous symmetry breaking of a gauge symmetry. As the temperature drops below a critical temperature $T_c$,  spontaneous symmetry breaking takes place. During inflation, the finite temperature effect instead comes from the Hawking-Gibbons temperature\cite{Gibbons:1977mu}, which is proportional to the Hubble parameter $H$. For simplicity, let us consider 
\begin{equation}
\label{varphipot}
V(\varphi) = (3H^2-M^2) |\varphi|^2 + \frac{\lambda |\varphi|^4}{4} + \frac{M^4}{\lambda}
\end{equation}
where $\varphi$ is a Higgs field in a specific representation of a symmetry. For example, $\varphi$ can belong to the adjoint  (the ${\underline {24}}$) representation of the $SU(5)$ grand unified gauge symmetry. The constant piece is introduced so that after spontaneous symmetry breaking and inflation, the resulting vacuum energy $V(\varphi)=0$.
Now, we can combine this with the potential (\ref{Xpot}), 
\begin{equation}
\label{Xvarphi}
V( X, \varphi)= \frac{1}{2} m^2 X^2 + V(\varphi) 
\end{equation}
Note that no term of the form $H^2 X^2$ is added because of the shift symmetry of $X$, which is broken only by a non-perturbative effect.
Spontaneous symmetry breaking of potential (\ref{Xvarphi}) starts at 
$$  3H^2= \frac{1}{2} m^2 X^2 + \frac{M^4}{\lambda} = M^2. $$
Before that the system is single field inflation with potential $\frac{1}{2} m^2 X^2 + \frac{M^4}{\lambda}$ and thereafter it becomes a two field inflation with $X$ and $\varphi$. The values of $M,\; \lambda$ are chosen such that the scale of inflation is correct and phase transition happens during inflation. Numerical study of this potential shows that for typical values of $M, \lambda$, the $n_s$ and $r$ values do not change significantly. For some extreme choice of $\lambda$, $<{\cal O}(10^{-8})$, $r$ will be pushed toward a smaller value since the constant $ \frac{M^4}{\lambda}$ in the potential lowers the value of $\epsilon$. 
Overall, we see that $r$ may be lowered slightly from the generic chaotic inflation value.

\section{Features of the Cosine Model}

We see that the helical model reduces to a single cosine model (\ref{Xcosine}),
\begin{equation}
\label{1cosine}
 V(\phi) = V_0\left\{ 1- \cos\left({\phi\over f}\right) \right\}
\end{equation} 
Although the single cosine model has been proposed before, a large value of $f$ is necessary but not always justified. Here, a value of $f \ge14  M_{pl}$ is fully justified in the helical model, as noted in Ref\cite{Kim:2004rp} some years ago.  Ref\cite{Kallosh:2014vja} shows that  a large class of axionic models reduce to the single cosine model. The periodicity of the potential is strongly indicative of axions. So it is reasonable to treat this single cosine model seriously.
Here we like to propose a simple way to search for the distinct feature of such a periodic potential. Since the $\phi^2$ chaotic inflation model is very simple and fits data quite well, we choose to study the cosine model by comparing with it. 

\subsection{Cosine Potential versus Chaotic Inflation}

Note that inflation ends when $\epsilon = {-\dot H/H^2}  > 1$ as $\phi$ decreases. Since there may be small terms that may be ignored during early stages of inflation, but may become important towards the end, we do not know the e-fold number $N_e$ (before the end of inflation) the CMB data covers, let us express all quantities in terms of the measurable quantities, namely, the tensor-scalar ratio $r$ and the power spectrum index $n_s$. In the chaotic inflation model, we have $r+4(n_s-1)=0$. Let us introduce $\Delta$ to measure the deviation from the $\phi^2$ chaotic inflation 
\begin{equation} 
\label{bigdelta}
          \Delta = {  r+ 4(n_s-1) \over 16 }.
\end{equation}
So $\Delta$ serves as a measure of the deviation from chaotic inflation since 
\begin{equation}
\label{signD}
  \mbox{chaotic inflation: } \Delta = 0 , \quad  \mbox{cosine model: }  \Delta = -{ 1\over 4f^2}.
\end{equation}

Eq.(\ref{signD}) shows that $\Delta$ is always negative; so, for fixed $n_s$, $r$ in the cosine model is always smaller than that in the $\phi^2$ model.
We present in Table 2 a comparison between the chaotic inflation and the cosine model in the slow-roll approximation,

\begin{center}
   \begin{tabular}{c| c c c c }
        $V(\phi) $  &&                                        ${1\over2} m^2 \phi^2$       &&  $V_0\left[ 1-\cos\left( {\phi \over f} \right) \right] $       \\  \hline   
         $\epsilon =  {1\over2}\left({V'\over V}\right)^2 $   &&      ${2\over \phi^2}$  &&        $     {1\over 2f^2} \cot^2 \left( {\phi\over 2f }\right)     $      \\
  $\eta= {V''\over V}$		&&      ${2\over \phi^2}$  &&        $     {1\over 2f^2} \left[  \cot^2\left( {\phi\over 2f }\right) -1  \right] $\\
  $\xi^2 = {V'V''' \over V^2}  $  		&&                  0             &&            $ -{1\over 2f^4} \cot^2 \left( {\phi\over 2f }\right) $  \\
  $ \omega^3 = {V'^2V''''\over V^3} $          &&             0             &&       $     {1\over 2f^6} \left[  \cot^4\left( {\phi\over 2f }\right) -\cot^2\left( {\phi\over 2f }\right)  \right] $ \\
  $n_s -1 = 2\eta - 6\epsilon $          &&       ${-8\over \phi^2}     $            &&                       $     -{1\over f^2} \left[  2\cot^2\left( {\phi\over 2f }\right) +1  \right] $   \\
   $r=16\epsilon$      &&              ${32\over \phi^2}      $        &&                  $ {8\over f^2} \cot^2 \left( {\phi\over 2f }\right)     $   \\

   \end{tabular}\\
 \end{center}

\begin{center} 
  {\small{} TABLE 2, Comparison between the $\phi^2$ chaotic inflation and the cosine model. We set  $M_{pl}=1$ in the above formulae, given in the slow-roll approximation. The notations are standard.}
\end{center}

Now we can express all quantities in terms of $\Delta$ and $r$ to show how the cosine model differs from chaotic inflation.
Following standard slow-roll inflation analysis, we include in TABLE 2, besides the slow-roll parameters, the scalar CMB spectrum spectral index $n_s$, its running  ${ dn_s \over d \ln k}$, the running of its running ${ d^2n_s \over d \ln k^2}$, the tensor spectrum spectral index $n_t$ and its running ${ dn_t \over d \ln k}$. Note that different experiments can choose different pivot wave number $k_*$.

\begin{center}
   \begin{tabular}{c| c c c c }
        $V(\phi) $  &&                                        ${1\over2} m^2 \phi^2$       &&  $V_0\left[ 1-\cos\left( {\phi \over f} \right) \right] $       \\  \hline   
         $\epsilon =  {1\over2}\left({V'\over V}\right)^2 $   &&      ${1\over 16}r$  &&        $   {1\over 16} r $      \\
  $\eta= {V''\over V} $		&&      ${1\over 16}r$  &&        $   {1\over 16}r + 2\Delta $\\
  $\xi^2 = {V'V''' \over V^2}  $  		&&                  0             &&            $ {1\over 2} r\Delta $  \\
  $ \omega^3 = {V'^2V''''\over V^3} $          &&             0             &&       $  {1\over32} r^2\Delta +r\Delta^2  $ \\
  $n_s -1 = 2\eta - 6\epsilon $          &&       $ -{1\over4}r  $            &&                       $  -{1\over4} r + 4\Delta $   \\
  $n_t =-2 \epsilon $ &&   $ -{1\over8}r  $ &&  $ -{1\over8}r  $ \\
$  { dn_s \over d \ln k} = -16\epsilon\eta + 24 \epsilon^2 +2\xi^2  $   &&        ${1\over 32} r^2 $        &&  $ {1\over 32} r^2 - r\Delta   $ \\
$  { dn_t \over d \ln k} = -4\epsilon\eta + 8 \epsilon^2   $   &&      ${1\over 64} r^2 $        &&  $ {1\over 64} r^2 -{1\over2} r\Delta  $ \\
 $ { d^2n_s \over d \ln k^2} = -192\epsilon^3 +192\epsilon^2 \eta -32\epsilon \eta^2 $ &&    $  -{1\over 128 } r^3 $   &&    $  -{1\over 128 } r^3 + {3\over 8}r^2 \Delta -4r \Delta^2 $     \\
 $ -24 \epsilon \xi^2 +2 \eta \xi^2 +2 \omega^3 $   && && \\
   \end{tabular}\\
 \end{center}

 \begin{center}
     {\small{} TABLE 3, Comparison between $\phi^2$ chaotic inflation and the cosine model for the various physically measurable quantities. $\Delta$ measures the difference between the 2 models. }
\end{center}

We see that all quantities in Table 3 can be expressed in terms of a single parameter, namely $r$ here, in the $\phi^2$ model, while the same quantities in the cosine model can be expressed in terms of 2 parameters, namely $r$ and $\Delta$ here (or equivalently, $r$ and $n_s$).  So there are a number of ways to extract the value of $\Delta$ when the CMB data improves. For example, besides Eq.(\ref{bigdelta}), one can extract the value of $\Delta$ just from the temperature power spectrum alone,

\begin{equation}
8\Delta^2 = \frac{1}{2} (n_s-1)^2 -  {dn_s \over d \ln k}
\end{equation}
 or just from the tensor perturbation data only,
\begin{equation}
\hat \Delta =16\Delta = {r\over 2} - {32\over r}{dn_t \over d \ln k}  \label{tenmeas}
\end{equation}
where we write the last equation in a way to remind ourselves that the shift of $r$ from the chaotic inflation value is actually ${\hat  \Delta}=16\Delta$. Here (\ref{tenmeas}) follow from (\ref{bigdelta}) within the slow-roll framework. Different measurements of $\Delta$ provide a test on the validity of the periodic potential in slow-roll inflation. Once the model is shown to be valid,  one can get a good measurement of $\Delta$ by combining different determinations of $\Delta$. 

\subsection{Bounds on $\hat \Delta$}

As $ f \rightarrow \infty$, the cosine model approaches chaotic inflation. However, $f$ is bounded from above.
Since $f \ge 14$, we see that the magnitude of $\Delta$ can be as big as  $|\Delta| \sim 10^{-3}$ while the CMB data indicates that the primordial $r$ coming from inflation is of order $10^{-1}$. So the downward shift of $r$, namely $\hat \Delta=16\Delta$, can be of order $10^{-2}$, or about as big as $10\%$ of $r$. 

To get an estimate of the range of $\hat \Delta$, recall that inflation together with CMB data implies $$m^2 =\frac{V_0}{f^2} \simeq 4 \times 10^{-11} M_{pl}^2$$ 
So we have, using the value of $V_{infl}$ (\ref{GUTinf}), 
\begin{equation}
 \hat \Delta =-\frac{4M_{pl}^2}{f^2} = -\frac{4m^2 M_{pl}^2}{V_0} = - 0.03 \frac{V_{infl}}{V_0} = - 0.03 \frac{(2 \times 10^{16} GeV)^4}{V_0}
 \end{equation}

We expect $V_0 \ge V_{infl}$, so the shift $\hat \Delta$ of $r$ should be no bigger than about 20\% of $r$. Table 3 shows that the runnings of the power spectrum indices in the cosine model can differ quite a bit from that in $\phi^2$ chaotic inflation.

The $r$ value shows that the inflationary scale $V_{infl}$ (\ref{GUTinf}) is very close to the GUT scale. To maintain the well-known coupling unification at the supersymmetric GUT scale, which is about $M_{GUT} \simeq 2 \times 10^{16}$ GeV, we should keep the string scale $M_S$ above it, so string effects do not mess up the coupling unification. For compactification, we like to keep $M_S$ smaller than $M_{pl}$. So
$$ M_{pl} \gg M_S \ge M_{GUT}$$ 
If $V_0$ is at the string scale with $M_S \simeq 10^{17}$ GeV, then $\hat \Delta$ is probably too small to be observed.

\subsection{Quartic Correction to the Chaotic Inflation Model}

We can expand the cosine model (\ref{1cosine}) to obtain
\begin{equation}
\label{phi4}
V(\phi)= \frac{1}{2} m^2 \phi^2\left( 1 - \frac{\phi^2}{12f^2} + . . .  \right) 
\end{equation}
Keeping only up to the quartic term, this $\phi^4$ corrected model yields, for fixed $\phi$,
\begin{equation}
\label{phifix}
\epsilon = \frac{2}{\phi_0^2} - \frac{1}{3f^2}, \quad \quad \eta =  \frac{2}{\phi_0^2} - \frac{5}{6f^2} 
\end{equation}
where we have chosen the same value of $\phi$, say at $\phi=\phi_0$ for both models, as in Table 2. The formulae for the linear (in $\Delta$) deviation in TABLE 3 apply equally well to this $\phi^4$ corrected model.

 \begin{center}
            \includegraphics[scale=0.4]{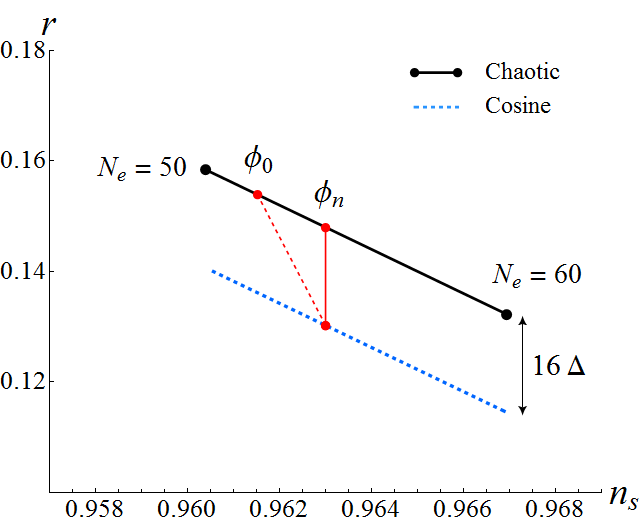} \\
            {\small{} FIG. 2  The black solid line is the prediction of $\phi^2$ chaotic inflation. The blue dotted line is the prediction of the cosine model. For  given values of $r$ and $n_s$, we can compare the 2 models with the same $\phi=\phi_0$ value (the red dashed line), or with the same $n_s$ value (the red solid line). In the latter case, $\phi_0$ in the cosine model now corresponds to $\phi_n$ in the chaotic inflation model.  The $16\Delta$ shown in the figure is about the maximum value one can get within the cosine model.}
 \end{center}

 Since the value of $n_s$ is relatively well measured, we may choose to fix its value instead when comparing the $\phi^4$ corrected model (\ref{phi4}) to chaotic inflation.
In this case, $\phi$ has to shift to 
$$\frac{2}{\phi_0^2} \rightarrow  \frac{2}{\phi_n^2} = \frac{2}{\phi_0^2} - \frac{1}{12f^2}$$
so $$\hat \epsilon = \frac{2}{\phi_n^2} -\frac{1}{4f^2} \quad \quad \hat  \eta =  \frac{2}{\phi_n^2} - \frac{3}{4f^2}$$ 
These two different choices of reference points are shown in FIG 2.
Physically, because the cosine potential is less steep for large $\phi$, we should start at a smaller of value of $\phi$, namely $\phi_0$  than the value, namely $\phi_n$, in the chaotic inflation to obtain the same value of $n_s$. That is, the same value of $n_s$ will translate to (slightly) different numbers of e-folds before the end of inflation.
As expected, 
this ambiguity disappears when we express all quantities in terms of physically measurable $r$, $n_s$ and $\Delta$.

In this form, our analysis can also be applied to other models such as the axion monodromy models \cite{Silverstein:2008sg,McAllister:2008hb,Marchesano:2014mla}. For large $M$ :
\begin{equation}
\label{monodromy}
V(\phi) = m^2M \left(\sqrt {M^2 + \phi^2} - M \right) \simeq \frac{1}{2} m^2 \phi^2\left( 1 - \frac{\phi^2}{4M^2} + . . .  \right) 
\end{equation}
In this case, the above analysis for the $\Delta$ to the linear order follows if we set $M^2 = 3f^2$.
For small $M$, the inflaton potential is closer to the linear form. In this case, it makes more sense to compare the axion monodromy model to the linear chaotic inflation model.

 For polynomial models, the $\phi^4$ term is more likely to enter with a positive sign. In that case, $\Delta >0$ and $r$ will be bigger than that given by chaotic inflation. 
However, such a positive $\phi^4$ term is absent in the cosine model. 
 
 Variations of the above model would introduce more parameters and so become less predictive. Nevertheless, that may be necessary if the data gets more precise. This will also mean that we can probe more detailed physics of the inflationary scenario.

\section{Cosmic Strings}

Here we like to argue that cosmic strings are quite natural to appear in any axionic models for inflation in string theory or SUSY grand unified theory \cite{Jeannerot:2003qv}. Not to confuse with the above axionic fields meant to drive inflation, we shall call them $a_i$ here. These axions differ from the above in that they have very small or no periodic potentials.

\subsection{Background}

Among the axion fields, some may have no periodic potential while others may couple to Abelian gauge fields. 
Consider a single axion field $a$ that couples to an $U(1)$ gauge field $A_{\mu}$.
Recall that the Lagrangian density will have a term of the form
$$ g^2 \rho^2(\partial_{\mu} a - A_{\mu})(\partial^{\mu} a - A^{\mu})/2$$
which is simply the mass term for the gauge field in the Abelian Higgs model after spontaneous symmetry breaking, with gauge coupling $g$ and vacuum expectation value $\rho$. Here, the gauge field $A_{\mu}$ swallows the axion to become massive in the Higgs mechanism. It is well known that this model has vortex solutions, or local strings. More generally, we can associate a string-like defect to every axion field. 
It is well-known that we may rewrite the axion field in terms of an anti-symmetric (2-form) field $B^{\lambda \kappa}$, where $ \epsilon_{\mu \nu \lambda \kappa} \partial^{\nu} B^{\lambda \kappa}=\partial_{\mu} a$. Here, the theory is invariant under a gauge transformation $B^{\lambda \kappa} \rightarrow B^{\lambda \kappa} + \partial^{\lambda} A^{\kappa} - \partial^{\kappa} A^{\lambda}$. Analogous to the gauge field case, where every 1-form field $A_{\mu}$ can have point particles charged under it, we can have strings (or vortices) charged under the 2-form field $B^{\lambda \kappa}$. 
It is interesting to note that the Higgs mechanism, in which the gauge field swallows the axion to become massive, may be re-interpreted as the $B^{\lambda \kappa}$ field swallows the gauge field to become a massive 2-form field.  

For an axion without coupling to a gauge field, it will have vortex solutions that are known as global strings. For an axion field with a small periodic potential of the form $T\cos (a/f)$, there are domain wall solutions. We expect to pass a domain wall as we go from $a \rightarrow a + 2\pi f$. Finite size domain walls are bounded by cosmic strings. As $T \rightarrow 0$, the tension of such a domain wall vanishes, leaving behind a closed string loop.  Typically there are a number of axions and $U(1)$ gauge fields in a string theory
mode. If the number of axions is larger than the number of $U(1)$ gauge fields, then some axions will lead to global strings. 

Cosmic strings produced at or after the end of inflation will eventually evolve to a scaling cosmic string network, so some of its overall properties can be reliably calculated. In general, the string tension dictates the rough overall properties of the cosmic string network, though details such as string tension spectrum, loop sizes and intercommutation probability are important too. 
The cosmological anisotropy generated by cosmic strings is qualitatively different from that from inflation. Inflation generates the metric and matter inhomogeneities which subsequently evolve unperturbed. Cosmic strings, on the other hand, actively generate scalar, vector and tensor perturbations throughout the history of the universe \cite{Hu:1997hp,Pen:1997ae,Turok:1997gj}. As the vector modes are completely negligible in the inflationary scenario since they quickly decay away, B-mode polarization in inflation comes entirely from the tensor perturbation. On the other hand, both vector and tensor perturbations continuously generated by a cosmic string network can contribute to the B-mode polarization\cite{Bevis:2007qz}. 

Temperature and density perturbations are actively being produced by cosmic strings, so the power spectrum does not have acoustic peaks. The presence of acoustic peaks in the power spectrum clearly rules out cosmic strings as the sole source of the temperature anisotropy. On the other hand, it is still possible that a small component of the power spectrum comes from the cosmic strings.  The CMB power spectrum puts a strong constraint on its possible contribution. Since the power spectrum generated by cosmic strings differs from that generated by inflation, one has to be more precise. The Planck Collaboration  \cite{Ade:2013xla} quantifies the amount of the anisotropy contributed by cosmic strings in terms of $f_{10}$, which is the fractional contribution of strings to the CMB temperature spectrum at multipole $\ell=10$, $f_{10} \equiv C^{\rm str}_{10} / C^{\rm tot}_{10}$. The first year Planck data constrains it at $f_{10} \lesssim 0.03$ \cite{Ade:2013xla}.  If there is only one type of cosmic strings with relatively large tension $\mu$, this 
corresponds roughly to $G\mu < 5 \times 10^{-7}$,  where $G= 1/8 \pi M_p^2$ is Newton's constant. 
 
\subsection{Constraints}
 
Since the scale of the inflaton potential is 
very close to the grand unified (GUT) scale, we expect the typical scale of the largest cosmic string tension to be $G \mu \sim 10^{-7}- 10^{-6}$.  This offers hope that CMB may detect the existence of the cosmic strings, since CMB measurements may reach $G \mu \ge 10^{-8}$. 
As discussed, B mode polarization may come from inflation, lensing which can converts E-mode to B-mode, and cosmic strings. Since they have different power spectrum, better data will be able to resolve them. Based on data from BICEP2 and POLARBEAR, 
recent analyses suggest the tantalizing possibility that adding some cosmic string contributions to the B-mode polarization is entirely acceptable, where the cosmic string contribution is close to but within the Planck bound $f_{10} < 0.03$ \cite{Lizarraga:2014eaa,Moss:2014cra}.

We note that pulsar timing actually has put a much stronger constraint on the cosmic string tension permitted than what we have allowed in the above discussion. Let us address this issue here. In pulsar timing arrays, gravitational wave can be detected through the modulation of arrival of pulses from millisecond pulsars due to the change in the distance between the earth and the pulsar caused by the gravitational wave bursts emitted by cosmic string loops from its cusps and kinks. The bound varies from $G \mu < 5 \times 10^{-7}$ to $G \mu < 10^{-10}$ \cite{Sanidas:2012ee}, depending on the details of the loop properties of the cosmic string network \cite{Vilenkin:book}.
This bound does not apply to global strings since they emit gravity waves at a much lower rate, due partly to the logarithmically divergent cores.  So the above bound applies only to local string network in the scaling limit. 

One way to avoid this pulsar timing bound is to delay the cosmic strings in reaching its scaling limit. If there were fewer cosmic strings produced during reheating, then it would take a longer time to reach its scaling limit. With fewer strings, one suppresses the gravitational wave bursts produced at early times and so avoids the pulsar timing bound. One way to achieve this has been suggested in Ref\cite{Kamada:2012ag}, where cosmic strings are produced during inflation. This way, cosmic string density can be diluted by inflation so there were very few cosmic strings right after inflation. Since the cosmic string tension and the energy scale of inflation are comparable, this possibility is rather reasonable; so here, we shall take a look at its features within the framework of the above model.

We can adapt the above model for this study. If we choose $\varphi$ in Eq.(\ref{varphipot}) and Eq.(\ref{Xvarphi}) to be a single complex scalar field, then its spontaneous symmetry breaking leads to the formation of vortices. If $\varphi$ couples to $U(1)$ gauge field, then we have local strings. Otherwise, we get global strings. Ref\cite{Kamada:2012ag} studies this case and concludes that strings can be diluted enough so that the pulsar timing bound can be avoided while CMB can still detect them if $G\mu > 7 \times 10^{-9}$.
If both monopoles and strings are produced during this time, monopole density goes like $1/a(t)^3$ while cosmic string density goes like $1/a(t)^2$ as the cosmic scale factor $a(t)$ grows after inflation. So the monopole density can be kept suppressed but the cosmic strings may come back to reach the scaling network limit, which scales like $1/a(t)^3$ in the matter dominated epoch. 

If there are more than one type of cosmic strings, it is not unreasonable that the production of heavier cosmic strings will be suppressed, in particular those close to the GUT scale, as discussed above. On the other hand, the production of cosmic strings with smaller tensions are probably produced after inflation and so should not be suppressed; that is, they can quickly reach their respective scaling limits.

\section{Conclusions}

We find that the helical model reproduces the (quadratic version) chaotic inflation model while the $r$ value may be a little smaller due to the periodic nature of the axion potential. We quantify this deviation, a distinctive feature of axionic inflation,  and argue that it can be tested and measured with better data. Phase transition during inflation will probably have little impact on $r$, or if anything, tends to lower its value. 

The data may allow some room for cosmic string contribution to the B-mode polarization. Since the B-mode power spectrum from cosmic strings differs substantially from that from inflation, better data offers the hope of detecting cosmic string signals or provides a tight bound on the possible cosmic string contribution.

We thank David Chernoff,  John Ellis, Mark Hindmarsh, Renata Kallosh, Andrei Linde, Levon Pogosian, Yoske Sumitomo, Alex Vilenkin and Ira Wasserman for discussions.

{\it While this paper was in preparation, Ref\cite{Choi:2014rja} by Choi, Kim and Yun appeared with the same helical model described here. After our paper, we note that Ref\cite{Kappl:2014lra,Ben-Dayan:2014zsa,Long:2014dta} also discuss the helical or a similar model. See also Ref\cite{Kaloper:2014zba,Harigaya:2014eta,Hebecker:2014eua,Higaki:2014pja,Bachlechner:2014hsa} for related proposals. The revised version of Ref\cite{Kallosh:2014vja} gives a simple nice realization of the helical model within supergravity. }


\begin{thebibliography}{99}

	\bibitem{Guth:1980zm} A.~H.~Guth,
  Phys.\ Rev.\ D {\bf 23}, 347 (1981).


	\bibitem{Linde:1981mu} A.~D.~Linde,
  Phys.\ Lett.\ B {\bf 108}, 389 (1982).

	\bibitem{Albrecht:1982wi} A.~Albrecht and P.~J.~Steinhardt,
  Phys.\ Rev.\ Lett.\  {\bf 48}, 1220 (1982).


\bibitem{Bennett:1996ce} C.~L.~Bennett, A.~Banday, K.~M.~Gorski, G.~Hinshaw, P.~Jackson, P.~Keegstra, A.~Kogut and G.~F.~Smoot {\it et al.},
  Astrophys.\ J.\  {\bf 464}, L1 (1996)
  [astro-ph/9601067].

	\bibitem{Bennett:2012zja} C.~L.~Bennett {\it et al.}  [WMAP Collaboration],
  Astrophys.\ J.\ Suppl.\  {\bf 208}, 20 (2013)
  [arXiv:1212.5225 [astro-ph.CO]].


	\bibitem{Ade:2013uln} P.~A.~R.~Ade {\it et al.}  [Planck Collaboration],
  arXiv:1303.5082 [astro-ph.CO].

  
	\bibitem{Starobinsky:1979ty} A.~A.~Starobinsky,
  JETP Lett.\  {\bf 30}, 682 (1979)
  [Pisma Zh.\ Eksp.\ Teor.\ Fiz.\  {\bf 30}, 719 (1979)].
  
 
	\bibitem{Crittenden:1993wm} R.~Crittenden, R.~L.~Davis and P.~J.~Steinhardt,
  Astrophys.\ J.\  {\bf 417}, L13 (1993)
  [astro-ph/9306027].


	\bibitem{Frewin:1993dq} R.~A.~Frewin, A.~G.~Polnarev and P.~Coles,
  Mon.\ Not.\ Roy.\ Astron.\ Soc.\  {\bf 266}, L21 (1994)
  [astro-ph/9310045].


	\bibitem{Harari:1993nb} D.~D.~Harari and M.~Zaldarriaga,
  Phys.\ Lett.\ B {\bf 319}, 96 (1993)
  [astro-ph/9311024].
  
	\bibitem{Kamionkowski:1996zd} M.~Kamionkowski, A.~Kosowsky and A.~Stebbins,
  Phys.\ Rev.\ Lett.\  {\bf 78}, 2058 (1997)
  [astro-ph/9609132].


	\bibitem{Seljak:1996gy} U.~Seljak and M.~Zaldarriaga,
  Phys.\ Rev.\ Lett.\  {\bf 78}, 2054 (1997)
  [astro-ph/9609169].


	\bibitem{Zaldarriaga:1996xe} M.~Zaldarriaga and U.~Seljak,
  Phys.\ Rev.\ D {\bf 55}, 1830 (1997)
  [astro-ph/9609170].


	\bibitem{Kamionkowski:1996ks} M.~Kamionkowski, A.~Kosowsky and A.~Stebbins,
  Phys.\ Rev.\ D {\bf 55}, 7368 (1997)
  [astro-ph/9611125].



	\bibitem{Ade:2014xna} P.~A.~R.~Ade {\it et al.}  [BICEP2 Collaboration],
  arXiv:1403.3985 [astro-ph.CO].
  

	\bibitem{Linde:1983gd} A.~D.~Linde,
  Phys.\ Lett.\ B {\bf 129}, 177 (1983).

\bibitem{Baumann:2014nda} 
  D.~Baumann and L.~McAllister,
  arXiv:1404.2601 [hep-th].


	\bibitem{Lyth:1996im} D.~H.~Lyth,
  Phys.\ Rev.\ Lett.\  {\bf 78}, 1861 (1997)
  [hep-ph/9606387].


	\bibitem{Freese:1990rb} K.~Freese, J.~A.~Frieman and A.~V.~Olinto,
  Phys.\ Rev.\ Lett.\  {\bf 65}, 3233 (1990).


	\bibitem{Adams:1992bn} F.~C.~Adams, J.~R.~Bond, K.~Freese, J.~A.~Frieman and A.~V.~Olinto,
  Phys.\ Rev.\ D {\bf 47}, 426 (1993)
  [hep-ph/9207245].


	\bibitem{Randall:1995dj} L.~Randall, M.~Soljacic and A.~H.~Guth,
  Nucl.\ Phys.\ B {\bf 472}, 377 (1996)
  [hep-ph/9512439].
  
\bibitem{Silverstein:2008sg} 
  E.~Silverstein and A.~Westphal,
  Phys.\ Rev.\ D {\bf 78}, 106003 (2008)
  [arXiv:0803.3085 [hep-th]].
  

	\bibitem{McAllister:2008hb} L.~McAllister, E.~Silverstein and A.~Westphal,
  Phys.\ Rev.\ D {\bf 82}, 046003 (2010)
  [arXiv:0808.0706 [hep-th]].
  
\bibitem{Marchesano:2014mla} 
  F.~Marchesano, G.~Shiu and A.~M.~Uranga,
  arXiv:1404.3040 [hep-th].
  
  \bibitem{Kim:2004rp} J.~E.~Kim, H.~P.~Nilles and M.~Peloso,
  JCAP {\bf 0501}, 005 (2005)
  [hep-ph/0409138].

\bibitem{Kaloper:2008fb} 
  N.~Kaloper and L.~Sorbo,
  Phys.\ Rev.\ Lett.\  {\bf 102}, 121301 (2009)
  [arXiv:0811.1989 [hep-th]].
  
\bibitem{Kallosh:2014vja} 
  R.~Kallosh, A.~Linde and B.~Vercnocke,
  arXiv:1404.6244 [hep-th].


	\bibitem{Ade:2014afa} P.~A.~R.~Ade {\it et al.}  [ The POLARBEAR Collaboration],
  arXiv:1403.2369 [astro-ph.CO].

	\bibitem{Lizarraga:2014eaa} J.~Lizarraga, J.~Urrestilla, D.~Daverio, M.~Hindmarsh, M.~Kunz and A.~R.~Liddle,
  arXiv:1403.4924 [astro-ph.CO].


	\bibitem{Moss:2014cra} A.~Moss and L.~Pogosian,
  arXiv:1403.6105 [astro-ph.CO].
  

\bibitem{Pogosian:2003mz} 
  L.~Pogosian, S.-H.~H.~Tye, I.~Wasserman and M.~Wyman,
  Phys.\ Rev.\ D {\bf 68}, 023506 (2003)
  [Erratum-ibid.\ D {\bf 73}, 089904 (2006)]
  [hep-th/0304188].
  

	\bibitem{Escoda:2003fa} C.~Escoda, M.~Gomez-Reino and F.~Quevedo,
  JHEP {\bf 0311}, 065 (2003)
  [hep-th/0307160].
 

	\bibitem{BlancoPillado:2004ns} J.~J.~Blanco-Pillado, C.~P.~Burgess, J.~M.~Cline, C.~Escoda, M.~Gomez-Reino, R.~Kallosh, A.~D.~Linde and F.~Quevedo,
  JHEP {\bf 0411}, 063 (2004)
  [hep-th/0406230].


	\bibitem{Kachru:2003aw} S.~Kachru, R.~Kallosh, A.~D.~Linde and S.~P.~Trivedi,
  Phys.\ Rev.\ D {\bf 68}, 046005 (2003)
  [hep-th/0301240].


	\bibitem{Gibbons:1977mu} G.~W.~Gibbons and S.~W.~Hawking,
  Phys.\ Rev.\ D {\bf 15}, 2738 (1977).
  
  
	\bibitem{Jeannerot:2003qv} R.~Jeannerot, J.~Rocher and M.~Sakellariadou,
  Phys.\ Rev.\ D {\bf 68}, 103514 (2003)
  [hep-ph/0308134].


	\bibitem{Hu:1997hp} W.~Hu and M.~J.~White,
  Phys.\ Rev.\ D {\bf 56}, 596 (1997)
  [astro-ph/9702170].

	\bibitem{Pen:1997ae} U.~-L.~Pen, U.~Seljak and N.~Turok,
  Phys.\ Rev.\ Lett.\  {\bf 79}, 1611 (1997)
  [astro-ph/9704165].


	\bibitem{Turok:1997gj} N.~Turok, U.~-L.~Pen and U.~Seljak,
  Phys.\ Rev.\ D {\bf 58}, 023506 (1998)
  [astro-ph/9706250].
  
	\bibitem{Bevis:2007qz} N.~Bevis, M.~Hindmarsh, M.~Kunz and J.~Urrestilla,
  Phys.\ Rev.\  D {\bf 76}, 043005 (2007)
  [arXiv:0704.3800 [astro-ph]].


	\bibitem{Ade:2013xla} P.~A.~R.~Ade {\it et al.}  [Planck Collaboration],
  arXiv:1303.5085 [astro-ph.CO].
  
\bibitem{Sanidas:2012ee} 
  S.~A.~Sanidas, R.~A.~Battye and B.~W.~Stappers,
  Phys.\ Rev.\ D {\bf 85}, 122003 (2012)
  [arXiv:1201.2419 [astro-ph.CO]].
  

\bibitem{Vilenkin:book} A.~Vilenkin and E.~P.~S.~Shellard, ``Cosmic Strings and Other Topological Defects'', Cambridge University Press (2000)


\bibitem{Kamada:2012ag} K.~Kamada, Y.~Miyamoto and J.~Yokoyama,
  JCAP {\bf 1210}, 023 (2012)
  [arXiv:1204.3237 [astro-ph.CO]].
   
 \bibitem{Choi:2014rja} K.~Choi, H.~Kim and S.~Yun,
  arXiv:1404.6209 [hep-th].
  
\bibitem{Kappl:2014lra} 
  R.~Kappl, S.~Krippendorf and H.~P.~Nilles,
  arXiv:1404.7127 [hep-th].
  
\bibitem{Ben-Dayan:2014zsa} 
  I.~Ben-Dayan, F.~G.~Pedro and A.~Westphal,
  arXiv:1404.7773 [hep-th].

\bibitem{Long:2014dta} 
  C.~Long, L.~McAllister and P.~McGuirk,
  arXiv:1404.7852 [hep-th].
  
\bibitem{Kaloper:2014zba} 
  N.~Kaloper and A.~Lawrence,
  arXiv:1404.2912 [hep-th].
  
\bibitem{Harigaya:2014eta}
K.~Harigaya and M.~Ibe,
arXiv:1404.3511 [hep-ph].

\bibitem{Hebecker:2014eua} 
  A.~Hebecker, S.~C.~Kraus and L.~T.~Witkowski,
  arXiv:1404.3711 [hep-th].
  
\bibitem{Higaki:2014pja} 
  T.~Higaki and F.~Takahashi,
  arXiv:1404.6923 [hep-th].
  
\bibitem{Bachlechner:2014hsa} 
  T.~C.~Bachlechner, M.~Dias, J.~Frazer and L.~McAllister,
  arXiv:1404.7496 [hep-th].


\end{thebibliography}
 \end{document}